\documentclass[american]{paper}
\usepackage[T1]{fontenc}
\usepackage[latin9]{inputenc}
\usepackage[a4paper]{geometry}
\usepackage{xcolor}
\usepackage{pdfcolmk}
\usepackage{babel}

\usepackage{prettyref}
\usepackage{amsmath}
\usepackage{graphicx}
\usepackage{amssymb}
\usepackage[authoryear]{natbib}
\PassOptionsToPackage{normalem}{ulem}
\usepackage{ulem}
\usepackage[unicode=true, 
 bookmarks=true,bookmarksnumbered=false,bookmarksopen=false,
 breaklinks=false,pdfborder={0 0 1},backref=false,colorlinks=false]
 {hyperref}
\hypersetup{pdftitle={The perfect integrator driven by Poisson input and its approximation in the diffusion limit},
 pdfauthor={Moritz Helias, Moritz Deger, Stefan Rotter, Markus Diesmann},
 pdfsubject={integrate-and-fire model},
 pdfkeywords={integrate-and-fire model, Fokker-Planck theory, diffusion},
 colorlinks=true,linkcolor=blue,citecolor=blue,urlcolor=blue,filecolor=blue}

\makeatletter

\providecolor{lyxadded}{rgb}{0,0,1}
\providecolor{lyxdeleted}{rgb}{1,0,0}

\newcommand{\lyxaddress}[1]{
\par {\raggedright #1
\vspace{1.4em}
\noindent\par}
}

\newrefformat{fig}{\hyperref[#1]{Fig.~\ref{#1}}}
\newrefformat{tab}{\hyperref[#1]{Table ~\ref{#1}}}
\newrefformat{sec}{\hyperref[#1]{Sec.~\ref{#1}}}
\newrefformat{sub}{\hyperref[#1]{Sec.~\ref{#1}}}
\newrefformat{alg}{\hyperref[#1]{Alg.~\ref{#1}}}
\newrefformat{app}{\hyperref[#1]{App.~\ref{#1}}}

\long\def\symbolfootnote[#1]#2{\begingroup%
\def\thefootnote{\fnsymbol{footnote}}\footnote[#1]{#2}\endgroup} 

\makeatother

\begin{document}
\global\long\def\ms{\:\mathrm{ms}}
\global\long\def\mV{\:\mathrm{mV}}
\global\long\def\Hz{\:\mathrm{Hz}}
\global\long\def\pA{\:\mathrm{pA}}
\global\long\def\pF{\:\mathrm{pF}}
\global\long\def\mus{\:\mu\mathrm{s}}
\global\long\def\Vth{V_{\theta}}
\global\long\def\Vr{V_{r}}
\global\long\def\PV{P(V)}
\global\long\def\Lfp{L_{\mathrm{FP}}}
\global\long\def\Qh{Q_{h}}
\global\long\def\Qp{Q_{p}}
\global\long\def\yr{y_{r}}
\global\long\def\yth{y_{\theta}}
\global\long\def\nuinst{\nu_{\mathrm{inst}}}
\global\long\def\cin{c_{\mathrm{in}}}
\global\long\def\defeq{\stackrel{\mathrm{def}}{=}}

\title{The perfect integrator driven by Poisson input and its approximation
in the diffusion limit}

\maketitle
\begin{center}
\textbf{\large Moritz Helias$^{1}$\symbolfootnote[1]{Corresponding author: Moritz Helias, \href{mailto:helias@brain.riken.jp}{helias@brain.riken.jp}},
Moritz Deger$^{2}$, Stefan Rotter$^{2,3}$, and Markus Diesmann$^{1,4,5}$ }
\par\end{center}{\large \par}

\lyxaddress{$^{1}$Computational Neurophysics, RIKEN Brain Science Institute,
Wako City, Japan\\
$^{2}$Bernstein Center Freiburg, Germany\\
$^{3}$Computational Neuroscience, Faculty of Biology, Albert-Ludwig
University, Freiburg, Germany\\
$^{4}$Institute of Neuroscience and Medicine, Computational and
Systems Neuroscience (INM-6), Research Center Juelich, Juelich, Germany\\
$^{5}$Brain and Neural Systems Team, Computational Science Research
Program, RIKEN, Wako City, Japan\\
}
\begin{abstract}
In this note we consider the perfect integrator driven by Poisson
process input. We derive its equilibrium and response properties and
contrast them to the approximations obtained by applying the diffusion
approximation. In particular, the probability density in the vicinity
of the threshold differs, which leads to altered response properties
of the system in equilibrium.
\end{abstract}

\subsection*{Stationary solution of perfect integrator with excitation}

The membrane potential $V$ of the perfect integrator \citep{Tuckwell88a}
evolves according to the stochastic differential equation \begin{align*}
\frac{dV}{dt} & =w\sum_{i}\delta(t-t_{i}),\end{align*}
where $t_{i}$ are random time points of synaptic impulses events
generated by a Poisson process with rate $\lambda$ and $w$ is the
magnitude is the voltage change caused by an incoming event. If $V$
reaches the threshold $\Vth$ the neuron emits an action potential.
After the threshold crossing, the voltage is reset to $V\leftarrow V-(\Vth-\Vr)$.
This reset preserves the overshoot above threshold and places the
system above the reset value by this amount. Biophysically the reset
is motivated by considering each $\delta$-impulse as the limit of
a current extended in time. If $V$ crosses $\Vth$ within such a
pulse, after the reset to $\Vr$ the remainder of the pulse's charge
causes a depolarization starting from $\Vr$. We consider a population
of identical neurons and assume a uniformly distributed membrane voltage
between reset and threshold initially. In what follows we apply the
formalism outlined in \citet{Helias10_1000929}. The first and second
infinitesimal moment \citep{Ricciardi99} of the diffusion approximation
are\begin{align*}
A_{1} & =\lambda w\defeq\mu\\
A_{2} & =\lambda w^{2}\defeq\sigma^{2}.\end{align*}
The corresponding neuron driven by Gaussian white noise hence obeys
the stochastic differential equation\[
\frac{dV}{dt}=\mu+\sigma\xi(t),\]
with the zero mean Gaussian white noise $\xi$, $\langle\xi(t)\xi(t+s)\rangle_{t}=\delta(s)$.
The probability flux operator is\[
S=\mu-\frac{\sigma^{2}}{2}\frac{\partial}{\partial V}.\]
We renormalize the stationary probability density $p(V)$ by the as
yet unknown flux $\nu$ as $q(V)=\frac{1}{\nu}p(V)$ so that the equilibrium
density fulfills the stationary Fokker-Planck equation\begin{equation}
Sq(V)=1_{V_{r}<V<V_{\theta}}.\label{eq:flux_eq}\end{equation}
Here $1_{\mathrm{expr.}}$ equals $1$ if $\mathrm{expr.}$ is true,
and $0$ else. The homogeneous solution of \prettyref{eq:flux_eq}
is\[
q_{h}(V)=e^{\frac{2\mu}{\sigma^{2}}V},\]
the particular solution which vanishes at $V=V_{\theta}$ for $V_{r}<V<V_{\theta}$
is\begin{align*}
q_{p}(V) & =-\frac{2}{\sigma^{2}}e^{\frac{2\mu}{\sigma^{2}}V}\int_{V_{\theta}}^{V}e^{-\frac{2\mu}{\sigma^{2}}u}\; du\\
 & =\frac{1}{\mu}\left(1-e^{\frac{2\mu}{\sigma^{2}}(V-V_{\theta})}\right).\end{align*}

We first consider the case of Gaussian white noise input of mean $\mu$
and variance $\sigma$. A finite probability flux in this case requires
$q(V_{\theta})=0$ at threshold. We hence obtain the full solution
that is continuous at reset as\[
q(V)=\frac{1}{\mu}\begin{cases}
1-e^{\frac{2\mu}{\sigma^{2}}(V-V_{\theta})} & \text{for }V_{r}<V<V_{\theta}\\
e^{\frac{2\mu}{\sigma^{2}}V}\left(e^{-\frac{2\mu}{\sigma^{2}}V_{r}}-e^{-\frac{2\mu}{\sigma^{2}}V_{\theta}}\right) & \text{for }-\infty<V<V_{r}.\end{cases}\]
The normalization $1=\int p(V)\; dV=\nu\int q(V)\; dV$ determines
the firing rate as\begin{equation}
\nu=\frac{\mu}{V_{\theta}-V_{r}}=\frac{\lambda w}{V_{\theta}-V_{r}},\label{eq:eql_rate}\end{equation}
With $\mu/\sigma^{2}=1/w$ the density is \begin{equation}
p(V)=\frac{1}{V_{\theta}-V_{r}}\begin{cases}
1-e^{\frac{2}{w}(V-V_{\theta})} & \text{for }V_{r}<V<V_{\theta}\\
e^{\frac{2}{w}V}\left(e^{-\frac{2}{w}V_{r}}-e^{-\frac{2}{w}V_{\theta}}\right) & \text{for }-\infty<V<V_{r}.\end{cases}\label{eq:p_V_gwn}\end{equation}
We next take into account the finite synaptic jumps to obtain a modified
boundary condition \citep{Helias10_1000929} at the firing threshold.
For $\Vr<V<\Vth$ the solution of \prettyref{eq:flux_eq} implies
a recurrence relation between higher derivatives, such that the $n$-th
derivative $q^{(n)}$ can be expressed in terms of the function value
itself as\begin{align*}
q^{\prime} & =\frac{2\mu}{\sigma^{2}}q-\frac{2}{\sigma^{2}}\\
q^{\prime\prime} & =\frac{2\mu}{\sigma^{2}}q^{\prime}=\frac{2\mu}{\sigma^{2}}\left(\frac{2\mu}{\sigma^{2}}q-\frac{2}{\sigma^{2}}\right)=\left(\frac{2\mu}{\sigma^{2}}\right)^{2}q-\mu\left(\frac{2}{\sigma^{2}}\right)\\
q^{(n)} & =\underbrace{\left(\frac{2\mu}{\sigma^{2}}\right)^{n}}_{d_{n}}q\underbrace{-\frac{2}{\sigma^{2}}\left(\frac{2\mu}{\sigma^{2}}\right)^{n-1}}_{c_{n}},\end{align*}
with $d_{0}=1$ and $c_{0}=0$ for completeness. Applying equation
(8) of \citet{Helias10_1000929} allows to determine the boundary
value at threshold as

\begin{align*}
q(V_{\theta}) & =\frac{1+\lambda\sum_{n=0}^{\infty}\frac{1}{(n+1)!}c_{n}(-w)^{n+1}}{-\lambda\sum_{n=0}^{\infty}d_{n}(-w)^{n+1}}\\
 & =\frac{1+\lambda\left(-\frac{1}{\sigma^{2}}w^{2}+\frac{1}{6}(-\frac{4\mu}{\sigma^{4}})(-w)^{3}\right)}{-\lambda\left(-w+\frac{\mu}{\sigma^{2}}w^{2}+\frac{1}{6}(2\frac{\mu}{\sigma^{2}})^{2}(-w^{3})\right)}\\
 & =\frac{\frac{4}{6}\frac{\mu}{\sigma^{4}}w^{3}}{\frac{4}{6}\frac{\mu^{2}}{\sigma^{4}}w^{3}}=\frac{1}{\mu}.\end{align*}
In the case of finite jumps, the region below reset will never be
entered, hence $q(V)=0$ for $V<V_{r}$. In order for the solution
to fulfill the boundary value at threshold the homogeneous solution
$q_{h}(V)=\mu^{-1}e^{2\mu(V-V_{\theta})/\sigma^{2}}$needs to be added
to the particular solution $q_{p}$, so the complete stationary density
is\[
q(V)=\begin{cases}
\frac{1}{\mu} & \text{for }V_{r}<V<V_{\theta}\\
0 & \text{for }-\infty<V<V_{r}.\end{cases}\]
The normalization therefore yields the same firing rate as in the
case of Gaussian white noise\begin{equation}
\nu=\frac{\mu}{V_{\theta}-V_{r}}=\frac{\lambda w}{V_{\theta}-V_{r}}.\label{eq:eql_rate_w}\end{equation}
This expression agrees with the intuitive expectation, because $\frac{\Vth-\Vr}{w}$
input impulses are needed to cause an output spike. Using this normalization,
the density is\begin{equation}
p(V)=\begin{cases}
\frac{1}{V_{\theta}-V_{r}} & \text{for }V_{r}<V<V_{\theta}\\
0 & \text{for }-\infty<V<V_{r}.\end{cases}\label{eq:p_V_w}\end{equation}

\begin{figure}

\centering{}\includegraphics{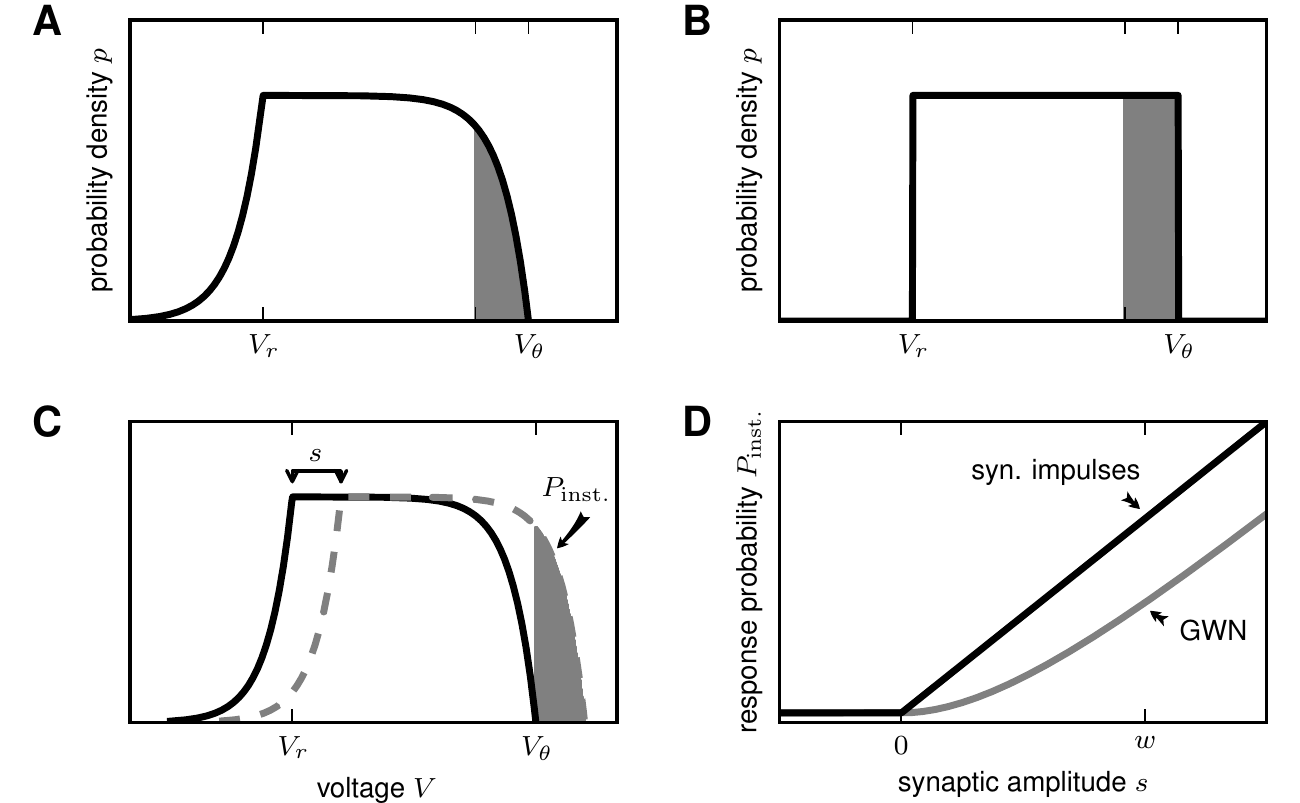}\caption{\textbf{Equilibrium density and response shaped by background.} (\textbf{A})
Probability density of voltage $V$ of a perfect integrator driven
by Gaussian white noise \prettyref{eq:p_V_gwn}. (\textbf{B}) Probability
density of a perfect integrator driven with excitatory synaptic impulses
of finite size $w$ causing the same drift and fluctuations as in
A \prettyref{eq:p_V_w}. The density near threshold most strongly
differs on the scale of the synaptic amplitude $w$ (gray shaded region).
(\textbf{C}) An additional excitatory impulse of amplitude $s$ shifts
the density (shown for Gaussian white noise background input as in
A), so that the gray shaded area exceeds the threshold. (\textbf{D})
The probability $P_{\text{inst.}}$ to respond with an action potential
corresponds to the area of density above threshold in C; it depends
on the shape of the density near threshold (black: background of synaptic
impulses of size $w$ \prettyref{eq:P_inst_w}; gray: Gaussian white
noise background \prettyref{eq:P_inst_gwn}). Parameters: $w=3\mV$,
$\Vr=0$, $\Vth=15\mV$ and $\lambda=200\;\frac{1}{s}$. \label{fig:voltage_densities}}

\end{figure}
The solutions for both cases are illustrated in \prettyref{fig:voltage_densities}A,B.

\subsection*{Instantaneous and time dependent response}

\begin{figure}

\begin{centering}
\includegraphics{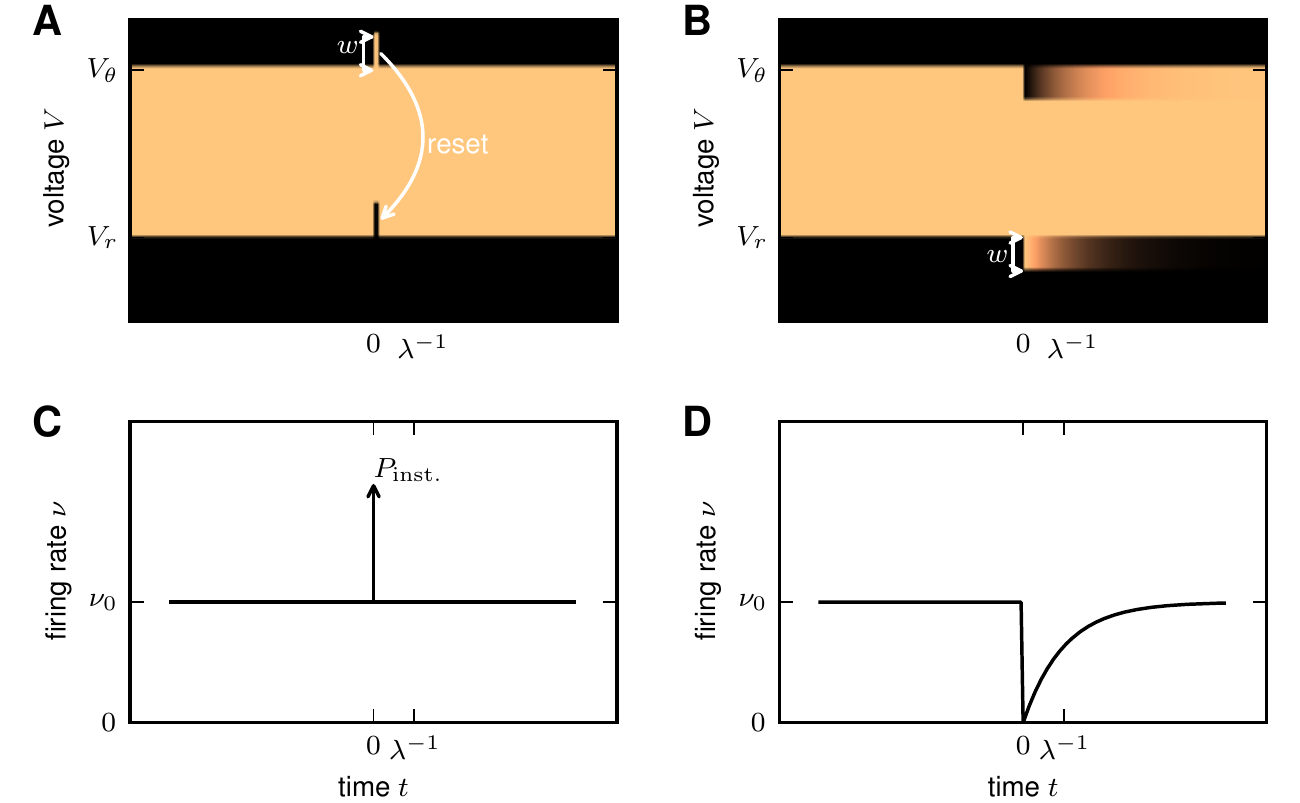}
\par\end{centering}

\caption{\textbf{Asymmetry of response.} (\textbf{A}) An additional excitatory
impulse of amplitude $w$ shifts the probability density upwards such
that a small part of the density exceeds the threshold. This leads
to an instantaneous spiking response, visible as a $\delta$-shaped
deflection in firing rate $\nu$ (visualized by bars of finite width
in\textbf{ A} and \textbf{C}). The reset of the membrane voltage to
$V_{r}$ after the spike moves the exceeding density down, so the
density immediately equals the state before the impulse. (\textbf{B})
An additional inhibitory impulse of amplitude $-w$ deflects the density
downwards \prettyref{eq:p_V_t_spike}. It does not cause a response
concentrated at the time of the impulse (\textbf{D}). Instead, the
firing rate $\nu$ instantaneously drops and exponentially reapproaches
its equilibrium value $\nu_{0}$ \prettyref{eq:nu_t_inh} as the density
gradually relaxes to its steady state on the time scale $1/\lambda$,
where $\lambda$ is the rate of synaptic background impulses.\label{fig:asymmetric_response}}

\end{figure}

The probability $P_{\text{inst.}}(s)$ that a neuron in the population
instantaneously emits an action potential in response to a single
synaptic input of postsynaptic amplitude $s$ equals the probability
mass $P_{\text{inst.}}(s)=\int_{V_{\theta}-s}^{V_{\theta}}p(V)\; dV$
crossing the threshold (shaded region in \prettyref{fig:voltage_densities}C).
In the the case of Gaussian white noise with \prettyref{eq:p_V_gwn}

\begin{align}
P_{\text{inst.}}(s) & =\int_{V_{\theta}-s}^{V_{\theta}}p(V)\; dV\nonumber \\
 & =1_{s>0}\left(\frac{s}{V_{\theta}-V_{r}}-\frac{w/2}{V_{\theta}-V_{r}}\left(1-e^{-\frac{2s}{w}}\right)\right)\nonumber \\
 & =1_{s>0}\frac{1}{V_{\theta}-V_{r}}\left(s+\frac{w}{2}\left(e^{-\frac{2s}{w}}-1\right)\right).\label{eq:P_inst_gwn}\end{align}
This expression grows quadratically like $P_{\text{inst.}}(s)\simeq1_{s>0}\frac{1}{V_{\theta}-V_{r}}\frac{s^{2}}{w}$
for small synaptic amplitudes $s$ as shown in \prettyref{fig:voltage_densities}D.
In the case of finite synaptic jumps using \prettyref{eq:p_V_w} we
get\begin{equation}
P_{\text{inst.}}(s)=1_{s>0}\frac{s}{V_{\theta}-V_{r}}.\label{eq:P_inst_w}\end{equation}
The response grows linear in the amplitude $s$ of the additional
perturbing spike (\prettyref{fig:voltage_densities}D). A linear approximation
of the integral response can be obtained using the slope of the equilibrium
rate \prettyref{eq:eql_rate_w} with respect to $\mu$ as\[
P_{\text{int.}}(s)=\int_{0}^{\infty}\nu(t)-\nu\; dt=s\frac{\partial\nu}{\partial\mu}=\frac{s}{V_{\theta}-V_{r}}.\]
For positive $s$ this expression equals the integral instantaneous
response \prettyref{eq:P_inst_w} so the complete response is instantaneous
in this case. For $s<0$ we only consider the special case of a synaptic
inhibitory pulse with the same magnitude $s=-w$ as the excitatory
background pulses, so the density is shifted away from threshold by
$w$ and the firing rate goes to $0$. The density reaches threshold
again if at least one excitatory pulse has arrived, which occurs within
time $t$ with probability $P_{k\ge1}=1-e^{-\lambda t}$. Given the
excitatory event, the hazard rate of the neuron is $\frac{\lambda w}{V_{\theta}-V_{r}}$,
so the time dependent response is\begin{equation}
\nu(t)=(1-1_{t>0}e^{-\lambda t})\frac{\lambda w}{V_{\theta}-V_{r}}.\label{eq:nu_t_inh}\end{equation}
The density after the inhibitory event therefore is a superposition
of the shifted density and the equilibrium density with the relative
weighting given by the probabilities $1-P_{k\ge1}$ and $P_{k\ge1}$,
respectively \begin{align}
p(V,t) & =\frac{1}{V_{\theta}-V_{r}}\begin{cases}
1_{t>0}e^{-\lambda t} & \text{for }V_{r}-w<V<V_{\theta}-w\\
1-1_{t>0}e^{-\lambda t} & \text{for }V_{r}<V<V_{\theta}.\end{cases}\label{eq:p_V_t_spike}\end{align}
The time evolution of the density following an excitatory and following
an inhibitory impulse at $t=0$ is shown in \prettyref{fig:asymmetric_response}
A and B, respectively.

The integrated response probability

\begin{align*}
P_{\text{int.}}(-w) & =\int_{0}^{\infty}\nu(t)-\nu\; dt\\
 & =-\frac{\lambda w}{V_{\theta}-V_{r}}\int_{0}^{\infty}e^{-\lambda t}\; dt\\
 & =-\frac{w}{V_{\theta}-V_{r}},\end{align*}
is the same as for an excitatory spike and coincides with the linear
approximation.

\subsection*{Stochastic resonance}

\begin{figure}

\begin{centering}
\includegraphics{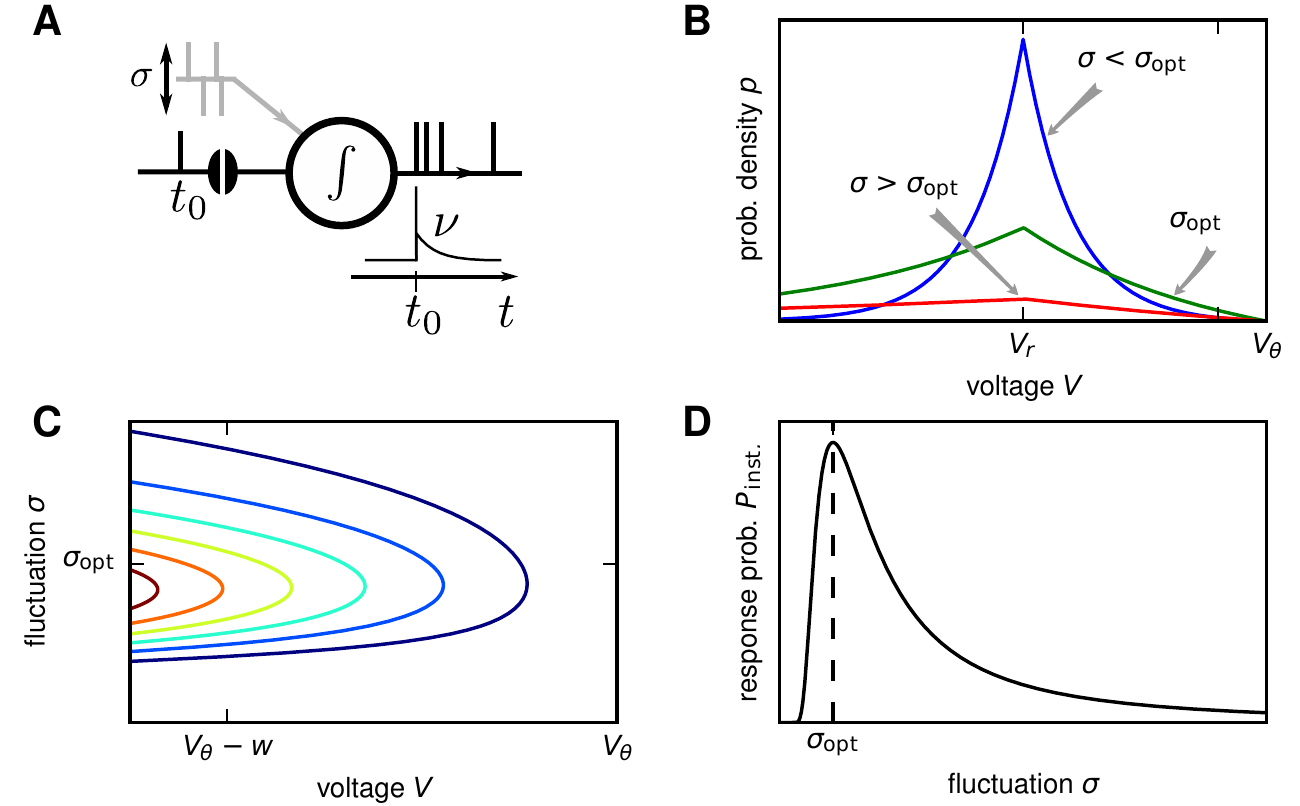}
\par\end{centering}

\caption{\textbf{Stochastic resonance.} (\textbf{A}) A model neuron receives
balanced excitatory and inhibitory background input (gray spikes).
The probability of a particular synaptic impulse (black vertical bar
at $t_{0}$) to elicit an immediate response by depends on the amplitude
$\sigma$ of the fluctuations caused by the other synaptic afferents.
(\textbf{B}) The spread of the probability density of voltage depends
on the amplitude $\sigma$ of the fluctuations caused by all synaptic
afferents \prettyref{eq:pdf_PI_const_leak}. At low fluctuations ($\sigma<\sigma_{\text{opt}}$)
it is unlikely to find the voltage near threshold, the density there
is negligible (blue: $\sigma=5.5\mV$). At intermediate fluctuations
($\sigma_{\text{opt}}$), the probability of finding the density below
threshold is elevated (green: $\sigma=11\mV$). Increasing the fluctuations
beyond this point ($\sigma>\sigma_{\text{opt}}$) spreads out the
density to negative voltages, effectively depleting the range near
threshold (red: $\sigma=16.5\mV$). (\textbf{C}) Zoom-in of the probability
density near threshold (luminance coded with iso-density lines) over
voltage $V$ (horizontal axis) as a function of the magnitude of fluctuations
$\sigma$ (vertical axis). At the optimal level $\sigma_{\text{opt}}$,
the density near threshold becomes maximal. (\textbf{D}) The voltage
integral of this density determines the probability of eliciting a
spike \prettyref{eq:P_inst} and has a single maximum at $\sigma_{\text{opt}}$.
Further parameters are $\Vr=0$, $\Vth=15\mV$, $\mu_{0}=5.0\mV/s$.\label{fig:stochastic_resonance}}

\end{figure}

In order to observe stochastic resonance, the fluctuation in the input
to the perfect integrator must be varied. We therefore consider a
zero mean Gaussian white noise input current $\sigma\xi(t)$. Adding
a constant restoring force $\mu(V)=-\mu_{0}\mathrm{sign}(V-V_{r})$,
$\mu_{0}>0$ assures that the voltage trajectories do not diverge
to $-\infty$ and approach $\Vr$ in absence of synaptic input. The
homogeneous solution of the stationary Fokker-Planck equation analog
to \prettyref{eq:flux_eq} therefore is $q_{h}(V)=e^{-\frac{2\mu_{0}}{\sigma^{2}}|V-V_{r}|}$.
The particular solution for $V>\Vr$ that fulfills the boundary condition
$q(V_{\theta})=0$ is found by variation of constants as

\[
q_{p}(V)=\frac{1}{\mu_{0}}\left(e^{-\frac{2\mu_{0}}{\sigma^{2}}(V-V_{\theta})}-1\right),\]
so the complete solution follows as\[
q(V)=\frac{1}{\mu_{0}}\begin{cases}
\left(e^{-\frac{2\mu_{0}}{\sigma^{2}}(V-V_{\theta})}-1\right) & \text{for }V_{r}<V<V_{\theta}\\
\left(e^{-\frac{2\mu_{0}}{\sigma^{2}}(V_{r}-V_{\theta})}-1\right)e^{\frac{2\mu_{0}}{\sigma^{2}}(V-V_{r})} & \text{for }-\infty<V<V_{r}.\end{cases}\]
Normalization again yields the equilibrium rate $\nu$

\begin{align*}
1 & =\nu\int q(V)\; dV\\
\nu & =\frac{\mu_{0}}{\frac{\sigma^{2}}{\mu_{0}}\left(e^{\frac{2\mu_{0}}{\sigma^{2}}(V_{\theta}-V_{r})}-1\right)+V_{r}-V_{\theta}},\end{align*}
and the normalized density is\begin{equation}
p(V)=\frac{\nu}{\mu_{0}}\begin{cases}
e^{\frac{2\mu_{0}}{\sigma^{2}}(V_{\theta}-V)}-1 & \text{for }V_{r}<V<V_{\theta}\\
\left(e^{\frac{2\mu_{0}}{\sigma^{2}}(V_{\theta}-V_{r})}-1\right)e^{\frac{2\mu_{0}}{\sigma^{2}}(V-V_{r})} & \text{for }-\infty<V<V_{r}.\end{cases}\label{eq:pdf_PI_const_leak}\end{equation}
\prettyref{fig:stochastic_resonance}B visualizes the density for
three different fluctuation amplitudes $\sigma$. In the limit of
large $\sigma^{2}\gg\mu_{0}$ the density decreases proportional to
$1/\sigma^{2}$ between reset and threshold and falls off linearly
towards threshold\[
p(V)\simeq\begin{cases}
\frac{2\mu_{0}}{\sigma^{2}}\frac{V_{\theta}-V}{V_{\theta}-V_{r}} & \text{for }V_{r}<V<V_{\theta}\\
\frac{2\mu_{0}}{\sigma^{2}}\left(1+\frac{2\mu_{0}}{\sigma^{2}}(V-V_{r})\right) & \text{for }-\infty<V<V_{r}.\end{cases}\]
The red curve in \prettyref{fig:stochastic_resonance}B shows the
tendency of such a linear decay towards threshold. The instantaneous
response exhibits stochastic resonance, because the integrated voltage
density near threshold assumes a maximum at a particular noise level
$\sigma$. This can already be judged from the zoom-in near threshold
in \prettyref{fig:stochastic_resonance}C. Formally, the response
to an incoming impulse of amplitude $s$ is\begin{align}
P_{\text{inst.}}(s) & =\int_{V_{\theta}-s}^{V_{\theta}}p(V)\; dV=\frac{1}{\frac{\sigma^{2}}{\mu_{0}}\left(e^{\frac{2\mu_{0}}{\sigma^{2}}(V_{\theta}-V_{r})}-1\right)+V_{r}-V_{\theta}}\left(-\frac{\sigma^{2}}{2\mu_{0}}\left(1-e^{\frac{2\mu_{0}}{\sigma^{2}}s}\right)-s\right)\nonumber \\
 & \stackrel{s\ll\sigma}{\simeq}\frac{1}{\frac{\sigma^{2}}{\mu_{0}}\left(e^{\frac{2\mu_{0}}{\sigma^{2}}(V_{\theta}-V_{r})}-1\right)+V_{r}-V_{\theta}}\;\frac{\mu_{0}}{\sigma^{2}}s^{2}\nonumber \\
 & =\frac{s^{2}}{\frac{\sigma^{4}}{\mu_{0}^{2}}\left(e^{\frac{2\mu_{0}}{\sigma^{2}}(V_{\theta}-V_{r})}-1\right)+\frac{\sigma^{2}}{\mu_{0}}\left(V_{r}-V_{\theta}\right)}.\label{eq:P_inst}\end{align}
The dependence on the noise is graphed in \prettyref{fig:stochastic_resonance}D.

\subsection*{\noindent Acknowledgements}

\noindent We acknowledge fruitful discussions with Carl van Vreeswijk,
Nicolas Brunel, Benjamin Lindner and Petr Lansky and thank our colleagues
in the NEST Initiative. Partially funded by BMBF Grant 01GQ0420 to
BCCN Freiburg, EU Grant 15879 (FACETS), EU Grant 269921 (BrainScaleS),
DIP F1.2, Helmholtz Alliance on Systems Biology (Germany), and Next-Generation
Supercomputer Project of MEXT (Japan).

\bibliographystyle{plainnat}
\bibliography{brain,math,computer}

\begin{thebibliography}{3}
\expandafter\ifx\csname natexlab\endcsname\relax\def\natexlab#1{#1}\fi
\expandafter\ifx\csname url\endcsname\relax
  \def\url#1{{\tt #1}}\fi

\bibitem[Helias et~al.(2010)Helias, Deger, Rotter, and
  Diesmann]{Helias10_1000929}
M.~Helias, M.~Deger, S.~Rotter, and M.~Diesmann.
\newblock Instantaneous non-linear processing by pulse-coupled threshold units.
\newblock {\em PLoS Comput Biol}, 6\penalty0 (9):\penalty0 e1000929, 2010.
\newblock doi:10.1371/journal.pcbi.1000929.

\bibitem[Ricciardi et~al.(1999)Ricciardi, Di~Crescenzo, Giorno, and
  Nobile]{Ricciardi99}
L.~M. Ricciardi, A.~Di~Crescenzo, V.~Giorno, and A.~G. Nobile.
\newblock An outline of theoretical and algorithmic approaches to first passage
  time problems with applications to biological modeling.
\newblock {\em MathJaponica}, 50\penalty0 (2):\penalty0 247--322, 1999.

\bibitem[Tuckwell(1988)]{Tuckwell88a}
Henry~C. Tuckwell.
\newblock {\em Introduction to Theoretical Neurobiology}, volume~1.
\newblock Cambridge University Press, Cambridge, 1988.
\newblock ISBN 0-521-35096-4.

\end{thebibliography}

\end{document}